\documentclass[12pt]{iopart}
\pdfoutput=1
\usepackage[italian, english]{babel}
\usepackage{amssymb,amsfonts,latexsym}
\usepackage[pdftex]{graphicx}

\begin{document}

\title{Off-equilibrium relaxational dynamics with improved Ising Hamiltonian}
\author{Mario Collura$^{1,2}$
}
\address{$^1$ 
Institut Jean Lamour, Department P2M, Groupe de Physique Statistique, Nancy-Universit\'{e} CNRS, B. P. 70239, F-54506 Vandoeuvre les Nancy Cedex, France}
\address{$^2$ Institut f\"{u}r Theoretische Physik, Universit\"{a}t Leipzig, Postfach 100 920, 04009 Leipzig, Germany}
\ead{mario.collura@ijl.nancy-universite.fr}

\begin{abstract}
We study the off-equilibrium  relaxational dynamics at criticality in the three-dimensional Blume-Capel model whose static critical behaviour belongs to the 3d-Ising universality class. Using ``improved'' Hamiltonian (the leading corrections to scaling have vanishing amplitude) we perform Monte Carlo simulations of the relaxational dynamics after a quench from $T=\infty$ to $T_c$. Analysing the off-equilibrium dynamics at $T_c$ we obtain an estimate of the dynamical critical exponent $z=2.020(8)$ that is perfectly consistent with the Field Theory predictions.
\end{abstract}

\section{Introduction}
Many statistical and dynamical properties of different physical systems show ``anomalies'' if the thermodynamics parameters approach a critical point. This singular behaviour exhibits a great universality. In fact, many properties of a system close to a continuos phase transition turn out to be largely independent of the microscopic details of the interactions between the individual atoms or molecules. They fall into one universality class characterized only by global features such as the symmetries of the underlying Hamiltonian, the number of spatial dimensions of the system, the number of components of order parameter and so on. Typically, close to a critical point, the correlation length and the other thermodynamics quantities exhibit power-law dependences on the parameters specifying the distance away from the critical point. These powers, or \textit{critical exponents}, are pure numbers, usually not integers or even simple rational numbers, which depend only on the universality class \cite{Ma, Cardy, ZinnJustin, PelissettoVicari}.

In critical dynamics we are interested in the universal features of the relaxational process: after preparing the system in a macroscopical initial state, we leave it to relax toward equilibrium. Upon approaching a critical point, the typical time scale of dynamics of the fluctuations around the equilibrium state diverges as $\sim\xi^z$ (\textit{critical slowing down}), where $\xi$ is the correlation length and $z$ is the dynamical critical exponent.

From theoretical point of view, after a critical quench, the dynamical behaviour far from equilibrium shows some universal features associated with the universal properties of the critical point. Thus, studying the off-equilibrium critical dynamics it is possible to acquire informations about the static and the dynamic universality class of the model and to get estimates of many different critical quantities.

So far, despite the efforts for understanding the off-equilibrium critical properties in many different systems \cite{Schepaul}, like two-dimensional Ising and Potts models \cite{Chatelain}, classical XY systems \cite{AbrietKarevski_2d, AbrietKarevski_3d}, disordered quantum spin chains \cite{AbrietKarevski_qs}, randomly diluted classical models \cite{relax_rand}, there is considerable theoretical and experimental interest to study some more fundamental systems, like 3d-Ising model, specially for having better numerical estimates of the critical exponents.

Indeed, some differences in the determination of the dynamical critical exponent $z$ for the three-dimensional Ising universality class yet exist between the numerical simulations, like Monte Carlo simulation at equilibrium \cite{Wansleben, Heuer} ($z=2.04(3)$, $z=2.10(2)$), damage spreading \cite{Matz,Grassberger,Gropengiesser} ($z=2.034(4)$, $z=2.04(1)$, $z=2.05(5)$), critical relaxation \cite{Stauffer&Knecht, Stauffer} ($z=2.05(2)$, $z=2.04(2)$), short-time dynamics \cite{Jaster} ($z=2.042(6)$), and the Field Theory predictions \cite{Bausch, Prudnikov} ($z\approx 2.02$, $z\approx 2.017$). The non-equilibrium relaxation method was also used to test the dynamic universality hypothesis in three-dimensional Ising models and two-dimensional three-states Potts models \cite{Murase}.

In order to have a better estimate of the dynamical critical exponent, we have studied a purely relaxational dynamics without conserved order parameter (also known as model A\footnote{The model A dynamical universality class can be specified in terms of the stochastic Langevin equation
\begin{equation*}
\partial_t\varphi(x,t)=-\Omega\frac{\delta\mathcal{H}[\varphi]}{\delta\varphi(x,t)}+\zeta(x,t),
\end{equation*}
where $\varphi(x,t)$ is the order parameter field, $\mathcal{H}[\varphi]$ is the static reduced Hamiltonian, $\Omega$ is a kinetic coefficient and $\zeta(x,t)$ is a zero-mean Gaussian white noise with correlations
\begin{equation*}
\langle\zeta(x,t)\zeta(y,s)\rangle=2\Omega\delta(x-y)\delta(t-s).
\end{equation*}
The critical dynamics of some anisotropic magnets and alloys are described by the previous dynamical equations with $\mathcal{H}$ given by the effective Hamiltonian of the Ising universality class.}) in 3d-Ising system; in particular, we have focused on the growth of correlations during the first instants of the evolution of the system after a quench from $T=\infty$ to $T=T_c$ (off-equilibrium critical dynamics). We have used a Blume-Capel Hamiltonian with improved parameters (see Sec. 2 for definitions and details). According to static and dynamics $\mathcal{RG}$ (Renormalization Group), the critical behaviour of such a system belongs into the Ising model universality class. For that reason, the dynamical critical exponent $z$ is the same. In addition, the leading scaling corrections appearing in dynamical quantities are governed by the same $\mathcal{RG}$ operators that control the non-asymptotic behaviour of static quantities and thus, are characterized by the same exponents as in the static case, i.e., by $\omega_{1}=0.84(4)$ and $\omega_{2} = 1.67(11)$ \cite{PelissettoVicari,Hasenbusch2,Newman}. As a consequence, the equilibrium improved Hamiltonians will cancel the leading scaling corrections also in the dynamical quantities. Therefore, the most precise estimates of dynamical universal quantities should be obtained in improved models, as in the static case \cite{Chen, Blote}.

We have been also careful at new scaling corrections introduced by dynamics. We have seen that the out-of-equilibrium dynamical quantities depend in general on the initial conditions and show analytical corrections tuned by $\tau_0 ^{-1}$ that controls the correlations of the order parameter at initial time surface.\footnote{If $\varphi(x,t)$ is the order parameter, then $\tau_{0}$ is defined via $\langle\varphi(x,t_{0})\varphi(y,t_{0})\rangle=\tau_{0}^{-1}\delta(x-y)$.} Using the dynamical $\mathcal{RG}$ methods one finds that it is possible to forget such as ``dynamical'' corrections in the data analysis.

We have performed MC (Monte Carlo) simulations for different lattice size $L$ and we have observed the initial temporal behaviour of susceptibility $\chi(t,L)$ and correlation length $\xi(t,L)$. We have noticed that for $L/\xi(t,L)\gtrsim5$ the finite-size effects are small and the behaviour of $\chi(t,L)$ is independent from lattice length. Using the susceptibility behaviour in this range, we have got our best value for the dynamical critical exponent
\begin{equation}
z=2.020(8).
\end{equation}
Our result is perfectly in agreement with FT (Field Theory) approach, both $\epsilon$-expansion \cite{Bausch} ($z\approx 2.02$) and fixed dimension \cite{Prudnikov} ($z\approx2.017$).

The paper is organized as follows. In Sec. 2 we define the models and the observables, giving also the scaling form predicted by the $\mathcal{RG}$ analysis. In Sec. 3 we report the analysis of MC simulations of the off-equilibrium relaxational critical behaviour in a quench from $T=\infty$ to $T_{c}$. Finally, we draw our conclusions in Sec. 4. Some details on the MC algorithm are discussed in Appendix A.

\section{Model and Observables}
We consider the Blume-Capel model with Hamiltonian
\begin{equation}
H=-\beta\sum_{\langle i,j\rangle} s_i s_j + D\sum_i s_i^2 , \label{blumecapel_MC}
\end{equation}
where the $s_i$ variables take the values  $\lbrace -1,0,+1\rbrace$, and the $i$ index runs over a cubic lattice with length $L$ and periodic boundary conditions. The brackets $\langle i,j\rangle$ indicate that sum runs over neighbors sites. $\beta$ and $D$ parameters are tuned at their improved critical values \cite{Hasenbusch,IHT,IHT2}:
\begin{equation}
D^{*} = 0.641,\qquad \beta_c = 0.3856717.
\end{equation}

According $D$ parameter at  the improved value $D^{*}$ minimizes the leading order scaling corrections and improves the dynamical behaviour also for small times. In short, using improved models we can obtain more reliable estimates about universal dynamical quantities. In fact, FSS (Finite Size Scaling) tells that, at criticality, in improved systems, the scaling corrections behave as $L^{-\omega_{2}}$ (with corresponding dynamic scaling corrections $t^{-\omega_{2}/z}$), where $\omega_{2}$ is the next to leading scaling corrections exponent; otherwise, in a generic system, the approach to the thermodynamics limit ($L\rightarrow\infty$) is slower, with scaling corrections that behave as $L^{-\omega_{1}}$ (with corresponding dynamical behaviour $t^{-\omega_{1}/z}$), with the leading scaling corrections exponent $\omega_{1}$.

In this improved model we have measured the Susceptibility
\begin{equation}
\chi(t)=\frac{1}{L^{3}}\bigg\langle\Big(\sum_{i} s_{i} (t) \Big)^2 \bigg\rangle , \label{chi(t)}
\end{equation}
and the Dynamics Structur Factor $C_{\vec{q}}(t)$ at least impulse ($|\vec{q}|=2\pi/L$)
\begin{equation}
C(t)=\frac{1}{3}\left\langle C_{\frac{2\pi}{L}\hat{x}}(t) +C_{\frac{2\pi}{L}\hat{y}}(t) + C_{\frac{2\pi}{L}\hat{z}}(t) \right\rangle \label{C(t)}
\end{equation}
with
\begin{equation}
C_{\vec{q}}(t)=\frac{1}{L^3}\bigg| \sum_{i}e^{\imath \vec{q}\cdot\vec{r}_{i}}s_{i}(t) \bigg|^2 ,
\end{equation}
where $\vec{r}_{i}\equiv (x_i,y_i,z_i)$ is the coordinate of the lattice site and $i$ runs over all the cubic lattice.

The system is invariant under axes permutation, so we have taken the average over the three spatial directions $(\hat{x},\hat{y},\hat{z})$ in order to improve the statistical average. By (\ref{chi(t)}) and (\ref{C(t)}) it is possible to define the Correlation Length $\xi$ using the discrete form
\begin{equation}
\xi(t)=\sqrt{\frac{\chi(t)/C(t) - 1}{4\sin^2(\pi/L)}. \label{corr_discreta}}
\end{equation}

\subsection{Susceptibility scaling form}
Using the dynamical $\mathcal{RG}$ methods and, taking into account further scaling corrections consequent by static properties of the Ising-like universality class, we expect that the magnetic susceptibility $\chi(t)$, in the infinite-volume limit, grows with MC time $t$ as \cite{relax_rand}
\begin{equation}
\chi(t)=\chi_0 t^{\rho}F_\chi (\mu_1 t^{-v_1},\mu_2 t^{-v_2},\ldots,\mu_n t^{-v_n}; \mu_{\tau_0} t^{-v_{\tau_{0}}}),
\end{equation}
where \cite{IHT}
\begin{equation}
\rho=\frac{2-\eta}{z},\qquad \eta = 0.0364(2). \label{ro_eta_def}
\end{equation} 
$F_\chi$ is analytic function of arguments. Scaling fields $\mu_i$ rule the statical scaling corrections, while $\mu_{\tau_0}$ is another scaling field linked to the off-equilibrium dynamics. Taking into account only the first two statical scaling fields and expanding in Taylor series, we obtain:
\begin{eqnarray}
\chi(t)&=& \chi_0 t^{\rho}\Big(1 + C_{11}t^{-v_1} + C_{12}t^{-2v_1} + \cdots + C_{21}t^{-v_2}+C_{22}t^{-2v_2} + \cdots \nonumber\\
&+& C_1^{(\tau_0)}t^{-v_{\tau_0}} + C_2^{(\tau_0)}t^{-2v_{\tau_0}} + \cdots \Big).\label{chi(t)_scaling}
\end{eqnarray}
According to the FT perturbative analysis, the leading scaling-corrections exponents should be the same as those that occur in equilibrium (static or dynamics) correlation functions. Therefore, we expect
\begin{equation}
v_1 = \frac{\omega_1}{z} \simeq 0.42(2), \qquad v_2 =\frac{\omega_2}{z}\simeq 0.83(5) ,
\end{equation}
where we have used $\omega_1=0.84(4)$, $\omega_2 = 1.67(11)\approx 2\omega_1$ and $z=2$ \cite{PelissettoVicari,Hasenbusch2,Newman}. The leading scaling correction proportional to $t^{-v_1}$ (and also all corrections of the form $t^{-kv_1}$) vanishes in improved models. Moreover, we discard the further scaling corrections due to $\tau_0^{-1}$ parameter because they are lower than corrections introduced by $v_2$ exponent, since $\omega_{\tau_{0}}=z$ and $v_{\tau_0}=\omega_{\tau_{0}}/z=1$.

Indeed, in the field-theoretical approach to non-equilibrium dynamics the expectation value of a generic observable has to be computed with a total dynamical functional given by $S[\varphi,\tilde{\varphi}] + \mathcal{H}_{0}[\varphi_{0}]$, where $S$ is the dynamical functional representation of the Langevin equation of the model A and $\mathcal{H}_{0}[\varphi_{0}]=\int d^d x \frac{\tau_{0}}{2}\varphi_{0}^2(x)$ provides the initial conditions. Then, following the standard methods of the dynamical $\mathcal{RG}$ \cite{Janssen, Calagamb}, the addition of $\mathcal{H}_{0}[\varphi_{0}]$ breaks the TTI (Time-Translational Invariance) and gives rise to new divergences in perturbation theory whenever time approaches the `time surface' located at $t_{0}=0$. It is possible to remove this new singularities without introducing new renormalization constants. Finally, it is possible to show that the $\mathcal{RG}$ dimension of $\mu_{\tau_{0}}$ is exactly $\omega_{\tau_0}\equiv z$. For that reason, we set henceforth $\mu_{\tau_0}=0$.

Equation (\ref{chi(t)_scaling}) is valid only in the infinite-volume limit. For a finite system of size $L$ we expect
\begin{equation}
\chi(t,L)=\chi_0 t^{\rho}E_0(tL^{-z})(1+C_{21}t^{-v_2}E_1(tL^{-z})+\cdots),\label{chi(t)_scaling2}
\end{equation}
where we have set the improved conditions and forgotten $\tau_0^{-1}$ corrections. Moreover, $E_n(x)$ are universal functions satisfying
\begin{eqnarray}
&&E_n(0)=1,\nonumber\\
&&E_0(x)\sim x^{-\rho},\quad E_{n\neq0}(x)\sim x^{n v_2}, \quad x\rightarrow\infty.
\end{eqnarray}

Even though, for improved models, from the theoretical point of view, it is perfectly correct to discard the leading-order scaling corrections, we should verify the absence of such a corrections in the data analysis. On the other hand, discarding corrections due to $\tau_0^{-1}$ should not affect the data analysis. Indeed, for our data range, the difference between $t^{-0.83}$ and $t^{-1}$ is small enough to allow us to absorb the effects of $\tau_0^{-1}$ into the corrections introduced by $v_2$ exponent (See Section 3 for more details).

\begin{figure}[t]
\includegraphics[width=0.5\textwidth]{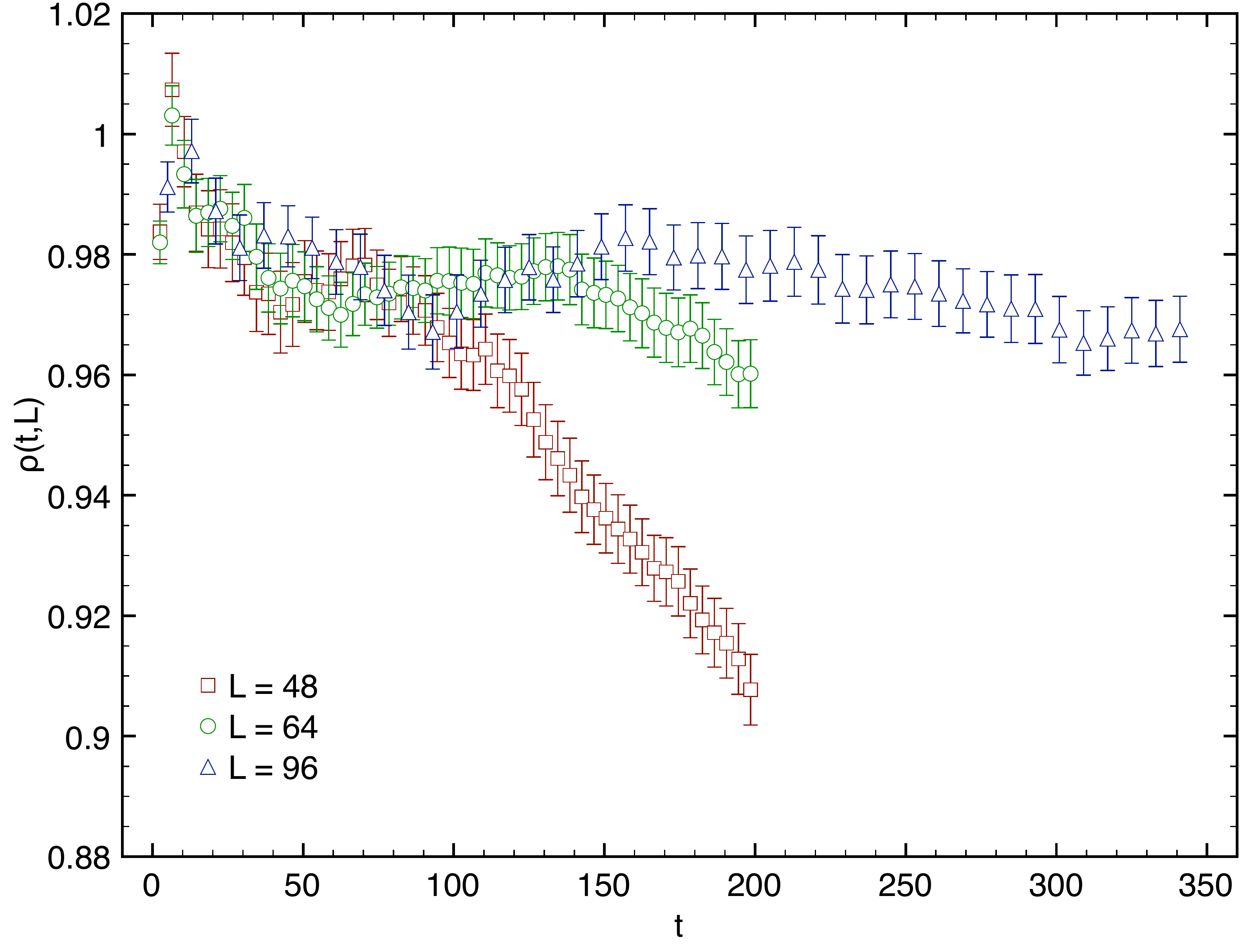}
\includegraphics[width=0.5\textwidth]{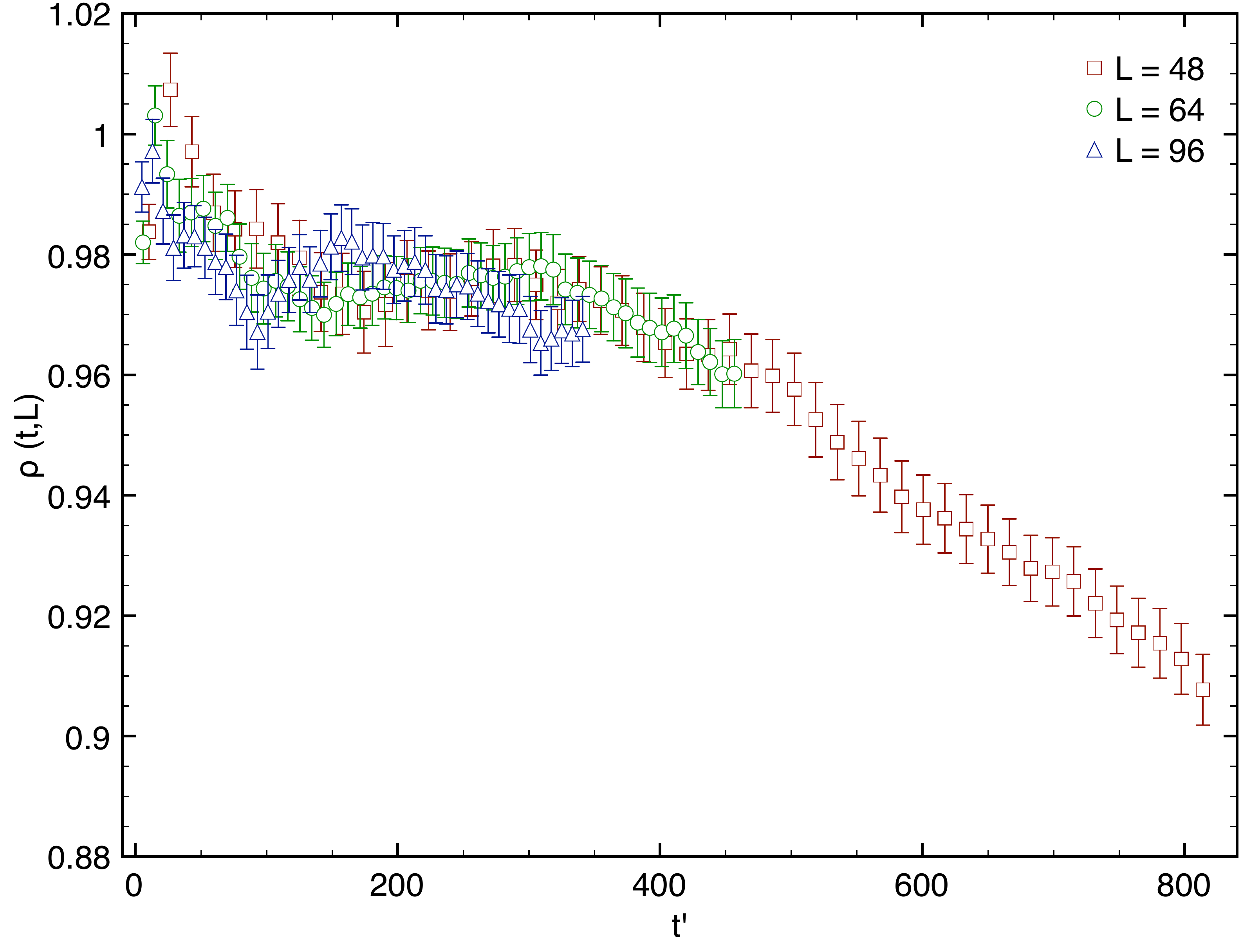}
\caption{\label{img_ro}\label{img_ror} On the left, the effective exponent $\rho_{eff}(t,L)$ for $L=48,64,96$. On the right, the same exponent versus the rescaled time $t'\equiv t(L/96)^{-z}$.}
\end{figure}

\section{Monte Carlo Simulation}
The off-equilibrium relaxational dynamics of the Ising model has already been investigated in \cite{Jaster}. Their result does not agree with the above-reported predictions. Reference \cite{Jaster} obtains $z=2.042(6)$. It is quite difficult to reconcile these results with the FT predictions; indeed, in \cite{Jaster} they do not take into account the corrections proportional to $t^{-0.42}$, i.e. controlled by the leading scaling-corrections exponent $\omega_1 = 0.84(4)$, therefore that result could be affected by this assumption. 

Others numerical MC works \cite{Stauffer&Knecht, Stauffer} investigate the critical relaxation starting from ordered states. Although their estimates of the dynamical critical exponent are still compatible within the error bar with the theoretical predictions (obtaining $z=2.05(2)$ and $z=2.04(2)$), we suppose that by using improved Hamiltonian we could obtain more reliable estimates of the dynamical critical exponent. 

Some effort was yet spent on the analysis of the 3d-Ising universality class by using improved Hamiltonian \cite{Chen, Blote}. Recently, the estimates of $D^{*}$ and $\beta_{c}$ for the Blume-Capel Hamiltonian parameters was improved, providing also new estimates for critical exponents $\nu$, $\eta$ and $\omega_{1}$ \cite{Hasenbusch3}. Thus, in the following, we further investigate this issue. We study the Metropolis dynamics of the Ising universality class after a quench from $T=\infty$ to $T_c$. This represents a nontrivial check of the FT predictions. 

\subsection{Estimate of the dynamic critical exponent $z$}
Since the model is improved, using only the next-to-leading scaling correction, we predict
\begin{equation}
\chi(t)= \chi_0 t^{\rho} (1+ C_{21} t^{-v_2}+\cdots). \label{chi_improved_scaling}
\end{equation}
We define an effective exponent
\begin{equation}
\rho_{eff}(t)\equiv \frac{\ln\left[ \chi(2t)/\chi(t)\right]}{\ln 2},\label{ro_eff_def}
\end{equation}
which, for $t\rightarrow\infty$, behaves as
\begin{equation}
\rho_{eff}(t)=\rho + C_{\rho} t^{-v_2} + \ldots,\qquad C_{\rho}=\frac{C_{21}(2^{-v_2}-1)}{\ln 2}.\label{ro_eff_scaling}
\end{equation}
On a finite lattice, (\ref{ro_eff_scaling}) is replaced by
\begin{equation}
\rho_{eff}(t,L)=\rho + e_0(t L^{-z}),\label{ro_eff_finitesize}
\end{equation}
where we have neglected large-$t$ corrections and $e_0(x)$ is a universal function (apart from a normalization of the argument) such that
\begin{equation}
e_0(0)=0, \quad \lim_{x\rightarrow\infty}e_0(x)=-\rho.
\end{equation}

We have performed off-equilibrium MC simulations on cubic lattices with periodic boundary conditions and size $L=48,64,96$ at $\beta=0.3856717$ and $D=0.641$ (our presently best estimate of $D^{*}$ is $D=0.641(8)$). For each lattice we have averaged over $2\cdot 10^5$ Markov chains. For each chain we have set different high-temperature initial conditions and, after a quench at critical point, we have done $400$ lattice sweeps ($700$ for $L=96$) using the Metropolis algorithm described in Appendix A.
\begin{figure}[t]
\includegraphics[width=0.5\textwidth]{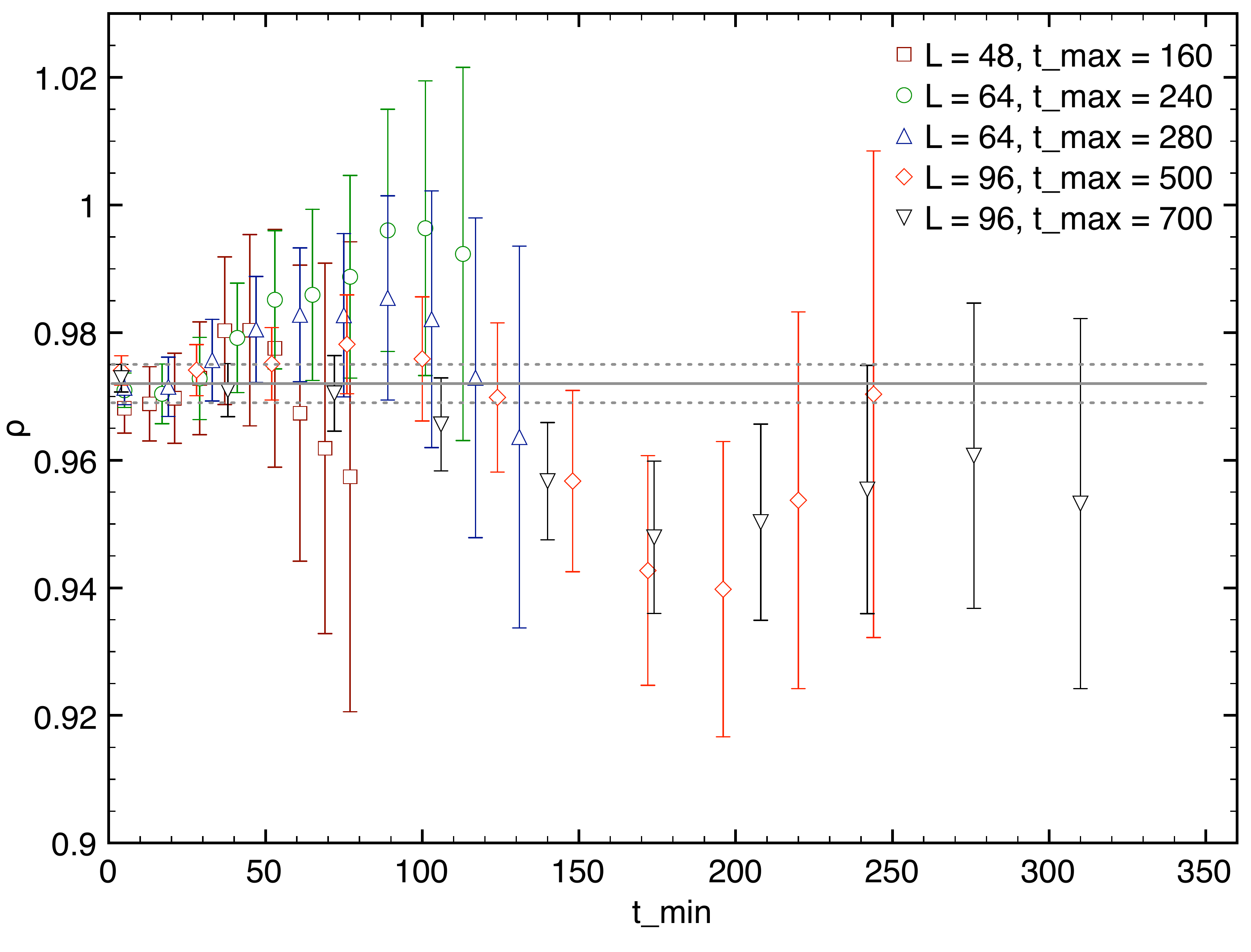}
\includegraphics[width=0.5\textwidth]{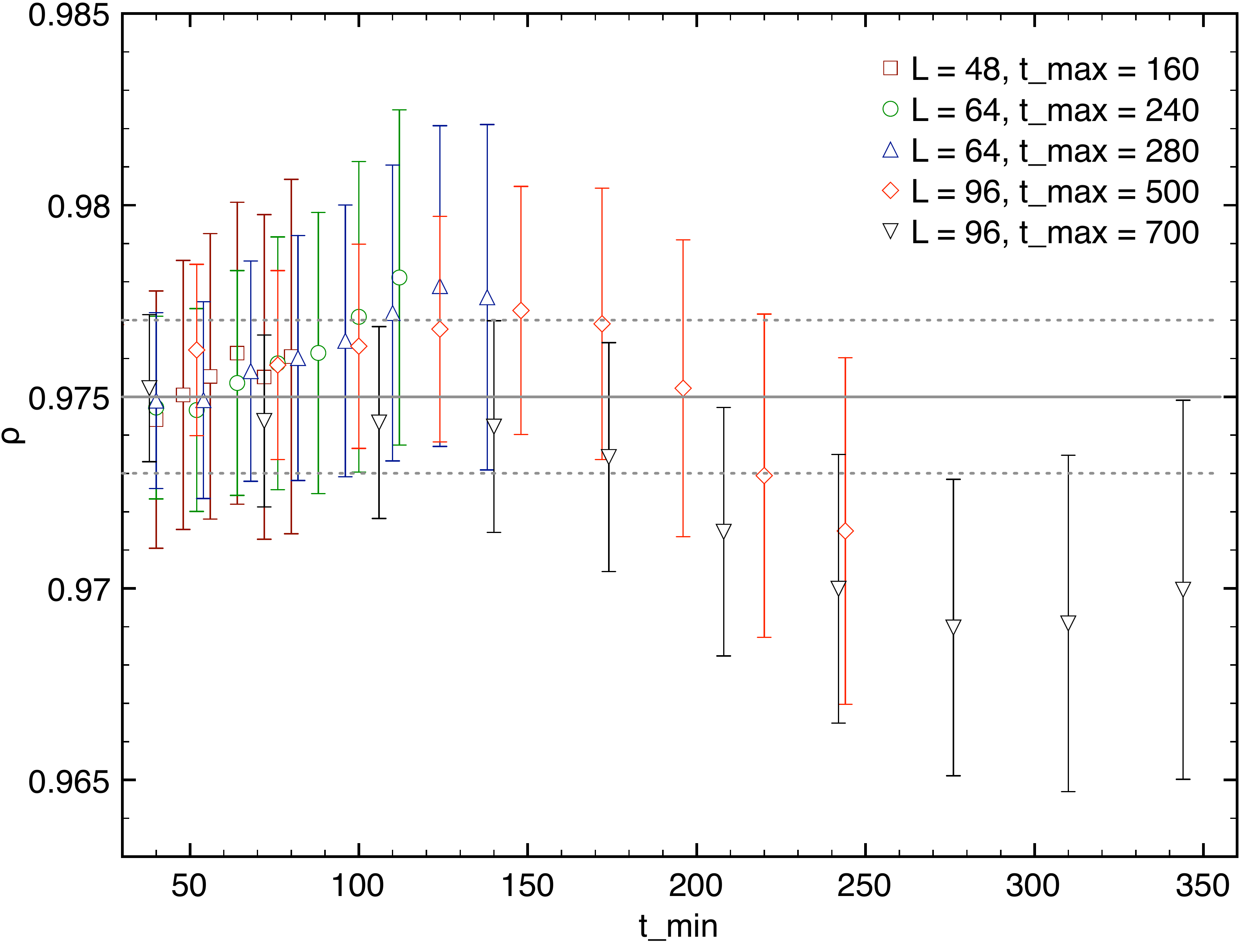}
\caption{\label{img_fit_scal}\label{img_fit_noscal}On the left, the results of fits $\ln\chi(t) = \ln\chi_{0} + \rho\ln t + C_{21}t^{-v_2}$ for $t_{min}\leqslant t \leqslant t_{max}$. The lines correspond to the estimate $\rho=0.972(3)$. On the right, the results of fits $\ln\chi(t) = \ln\chi_{0} + \rho\ln t$ for data that do not show corrections to scaling. The lines correspond to the estimate $\rho=0.975(2)$. Some data are slightly shifted along the $x$ axis to make them visible.}
\end{figure}

In Fig.\ref{img_ro} (on the left) we show $\rho_{eff}(t,L)$ for $L=48,64,96$. It clearly shows finite-size corrections, and, for each $L$, $\rho_{eff}(t,L)$ follows the infinite-volume curve up to an $L$-dependent value $t_{max}(L)$. As shown by Fig.\ref{img_ror} (on the right), where $\rho_{eff}(t,L)$ is plotted versus $t_{resc}=t(L/96)^{-z}$, finite-size effects are consistent with (\ref{ro_eff_finitesize}). Thus, the value $t_{max}(L)$, after which finite-size effects cannot be neglected, increases as $L^z$. Infinite-volume quantities, such as $\rho_{eff}(t)$, must be obtained from the data at $t<t_{max}(L)$. Fig.\ref{img_ro} indicates that, with the statistical errors of our data, $t_{max}(L)\approx 90,150$ for $L\approx 48,64$; moreover, using FSS Ansatz, we also predict $t_{max}(L=96)\approx350$.\footnote{In Fig.\ref{img_ror} we have used our estimate for the dynamical critical exponent $z$. Instead, for the scaling analysis of $t_{max}$ we have used the mean field approximation value $z=2$.}

Since $\rho_{eff}(t,L)$ is defined using data at $t$ and $2t$, this implies that, for $L=96$, only data corresponding to $t\lesssim 700$ have negligible finite-size effects within our error bars. Finite-size effects give rise to  systematic error in the estimate of $\rho$. As is clear from Fig.\ref{img_ro}, they yield smaller values of $\rho$, and therefore larger values of $z$.

After selecting the temporal range without finite-size effects, from the susceptibility behaviour in this range, we have achieved our estimate of the dynamical critical exponent $z$. We have fitted the data by using (\ref{chi_improved_scaling}): we have defined the temporal range $[t_{min},t_{max}]$ with $t_{max}(L=48)=160$, $t_{max}(L=64)=240,280$ and $t_{max}(L=96)=500,700$; for each $t_{max}$, $t_{min}$ was moved in the interval $[5,t_{max}/2]$. In this way we have checked the stability of our result.
\begin{figure}[t]
\center
\includegraphics[width=0.5\textwidth]{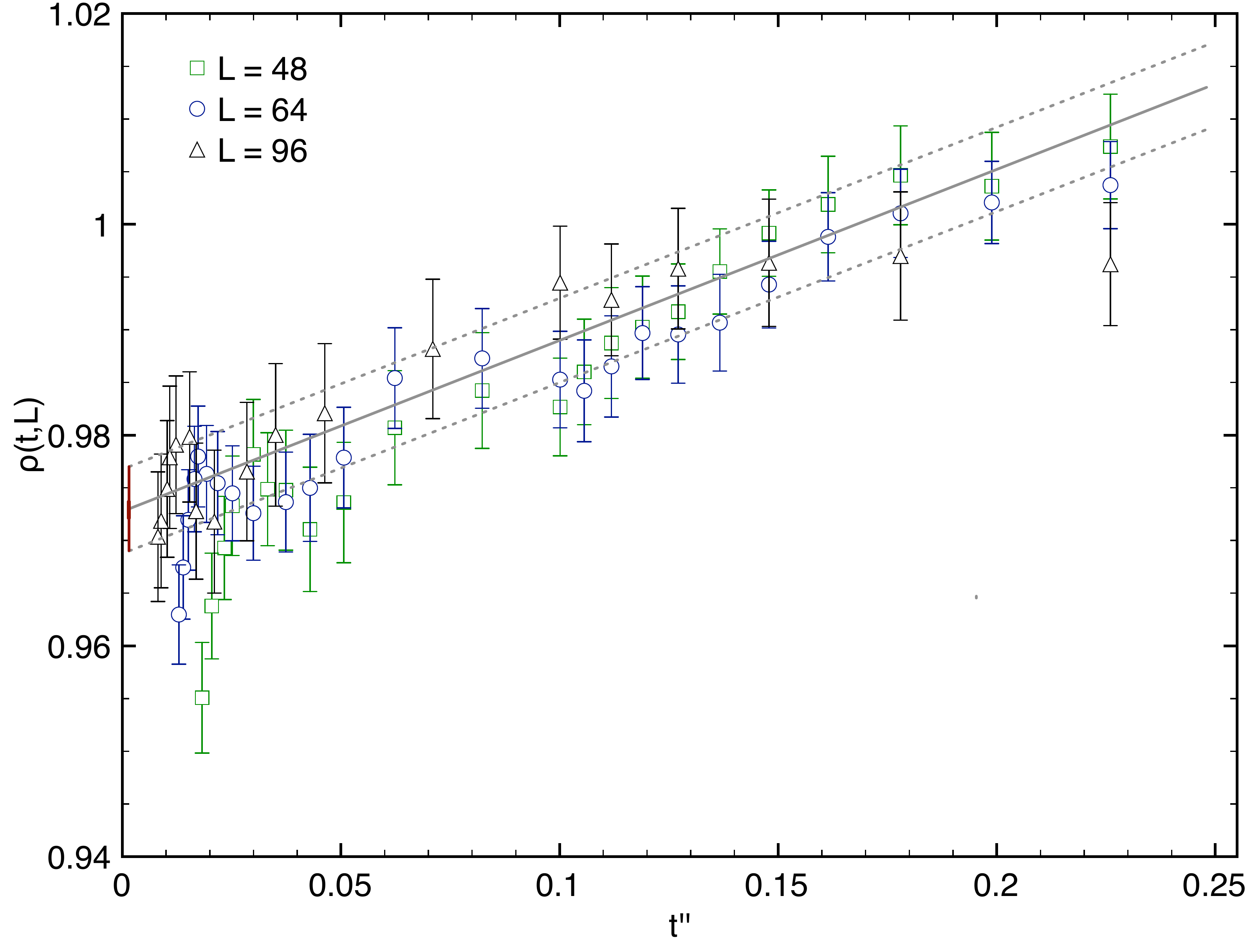}
\caption{\label{img_ror2}The effective exponent $\rho_{eff}(t,L)$ versus $t''\equiv t^{-v_2}$ with $v_2=0.83$. The lines are drawn to guide the eyes.}
\end{figure}

In Fig.\ref{img_fit_scal} (on the left) we plot the results of fits of $\ln\chi(L,t)$ for $t_{min}\leqslant t \leqslant t_{max}$  to $\ln\chi_{0} + \rho\ln t + C_{21}t^{-v_2}$ with $v_2 = 0.83$, for $L=48,64,96$. Finally, in Tab.\ref{tabella_ro_scal} we report our best estimate of $\rho$ for different $L$ and $t_{max}$ getting into account the next-to-leading scaling corrections.
\begin{table}[t]
\begin{center}
\begin{tabular}{@{\hspace{0.5cm}}l@{\hspace{0.5cm}}l@{\hspace{0.5cm}}l@{\hspace{0.5cm}}l@{\hspace{0.5cm}}l@{\hspace{0.5cm}}}
\hline\hline
 $t_{max}$& $t_{min}$ & $L=48$ & $L=64$ & $L=96$ \\
\hline
 & $20$ &  $0.970(7)$ & - & - \\
 & $40$ &  $0.981(13)$ & - & - \\
$160$ & $60$ &  $0.97(2)$ & - & - \\
 & $80$ &  $0.94(4)$ & - & - \\
  & $100$ &  $0.94(6)$ & - & - \\
\hline
 & $20$ & - & $0.970(5)$ & - \\
 & $40$ & - & $0.979(8)$ & - \\
$240$ & $60$ & - & $0.986(12)$ & - \\
 & $80$ & - & $0.990(17)$ & - \\
  & $100$ & - & $0.99(2)$ & - \\
\hline
 & $20$ & - & $0.972(5)$ & - \\
 & $40$ & - & $0.978(7)$ & - \\
$280$ & $60$ & - & $0.983(10)$ & - \\
 & $80$ & - & $0.983(14)$ & - \\
  & $100$ & - & $0.983(19)$ & - \\
\hline
 & $20$ & - & - & $0.974(4)$ \\
  & $40$ & - & - & $0.975(5)$ \\
$500$ & $60$ & - & - & $0.976(6)$ \\
 & $80$ & - & - & $0.978(8)$ \\
  & $100$ & - & - & $0.976(9)$ \\
\hline
  & $20$ & - & - & $0.972(3)$ \\
  & $40$ & - & - & $0.971(4)$ \\
$700$ & $60$ & - & - & $0.971(5)$ \\
 & $80$ & - & - & $0.970(6)$ \\
  & $100$ & - & - & $0.968(7)$ \\
\hline\hline
\end{tabular}
\caption{\label{tabella_ro_scal}Some best estimate of $\rho$ for different $L$, $t_{max}$ and $t_{min}$.}
\end{center}
\end{table}
Data corresponding to different lattice sizes are not correlated. Moreover, there are not finite-size effects in the data, thus we have got the estimate of the $\rho$ exponent (in the thermodynamic limit $L\rightarrow\infty$):
\begin{equation}
\rho = 0.972(3), \label{ro_stima_scal}
\end{equation}
and, using (\ref{ro_eta_def}), we obtain
\begin{equation}
z=2.020(6). \label{z_stima_scal}
\end{equation}

Finally, in Fig.\ref{img_ror2} we plot $\rho_{eff}(t,L)$ for $L=48,64,96$ versus $t^{-v_2}$ with $v_2=0.83$. Finite-size effects are negligible for $t^{-v_2}>t_{max}(L)^{-v_2}\approx 0.015, 0.009, 0.004$, for $L=48,64,96$ respectively. The data satisfying this inequality clearly follow a unique curve, which is expected to behave as $\rho + ct^{-v_2}$ for sufficiently large values of $t$. The data plotted in Fig.\ref{img_ror2} clearly show such a behaviour in the region $t^{-v_2}\lesssim 0.2$ (corresponding to $t\gtrsim 6$), and are perfectly compatible with the value $\rho=0.972(3)$.

\paragraph*{Result without corrections to scaling -}
After that, we have also achieved an estimate of the exponent $z$ by using data range that should not show next-to-leading scaling corrections. Thus, we have fitted the data with $\ln\chi(t) = \ln\chi_{0} + \rho\ln t$. The results are shown in Tab.\ref{tabella_ro_noscal} and in Fig.\ref{img_fit_noscal} (on the right).

\begin{table}[t]
\begin{center}
\begin{tabular}{@{\hspace{0.5cm}}l@{\hspace{0.5cm}}l@{\hspace{0.5cm}}l@{\hspace{0.5cm}}l@{\hspace{0.5cm}}l@{\hspace{0.5cm}}}
\hline\hline
 $t_{max}$& $t_{min}$ & $L=48$ & $L=64$ & $L=96$ \\
\hline
 & $40$ &  $0.974(3)$ & - & - \\
$160$ & $60$ &  $0.976(4)$ & - & - \\
 & $80$ &  $0.976(5)$ & - & - \\
  & $100$ &  $0.973(6)$ & - & - \\
\hline
 & $40$ & - & $0.975(2)$ & - \\
$240$ & $60$ & - & $0.975(3)$ & - \\
 & $80$ & - & $0.976(3)$ & - \\
  & $100$ & - & $0.977(4)$ & - \\
\hline
 & $40$ & - & $0.975(2)$ & - \\
$280$ & $60$ & - & $0.975(3)$ & - \\
 & $80$ & - & $0.976(3)$ & - \\
  & $100$ & - & $0.977(4)$ & - \\
\hline
  & $40$ & - & - & $0.976(2)$ \\
$500$ & $60$ & - & - & $0.976(2)$ \\
 & $80$ & - & - & $0.976(2)$ \\
  & $100$ & - & - & $0.976(3)$ \\
\hline
  & $40$ & - & - & $0.975(2)$ \\
$700$ & $60$ & - & - & $0.975(2)$ \\
 & $80$ & - & - & $0.974(2)$ \\
  & $100$ & - & - & $0.974(2)$ \\
\hline\hline
\end{tabular}
\caption{\label{tabella_ro_noscal}Some best estimate of $\rho$ for different $L$, $t_{max}$ and $t_{min}$. Here we only consider data that do not show corrections to scaling.}
\end{center}
\end{table}

Also in this case, data corresponding to different lattice sizes are not correlated and there are not finite-size effects. We have got the estimate
\begin{equation}
\rho = 0.975(2),\quad z=2.014(4),  \label{z_stima_noscal}
\end{equation}
perfectly consistent with (\ref{z_stima_scal}).

\paragraph*{Errors analysis - }
As already mentioned at the end of Section 2, we can discard the corrections due to $\tau_0^{-1}$ ($\sim t^{-1}$) since they are of the same order of those proportional to $t^{-0.83}$. We have seen, in fact, that fitting the data adding a new term proportional to $t^{-1}$ did not change the results and introduced instability to the fit. It is easy to see that for $t\in[20,500]$ one has $t^{-0.83}\in[0.08,0.006]$ and $t^{-1}\in[0.05,0.002]$, thus $t^{-1}\approx t^{-0.83}$. As a consequence, the effects of $t^{-1}$ are absorbed by $t^{-0.83}$.

\begin{figure}[t]
\center
\includegraphics[width=0.5\textwidth]{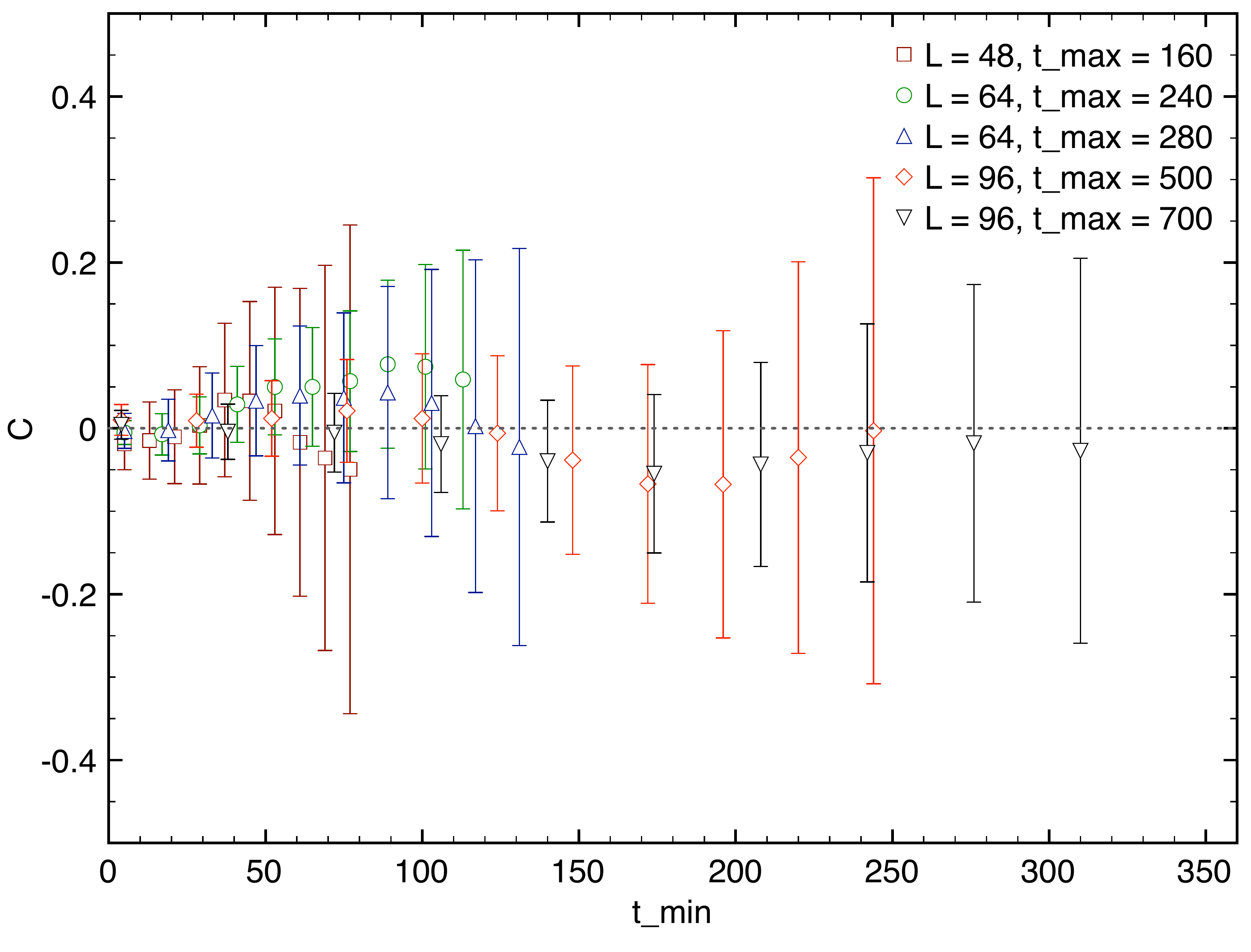}
\caption{\label{img_C} Coefficient $C_{11}$ for the fit $\ln\chi(t) = \ln\chi_{0} + \rho\ln t + C_{11}t^{-v_1} + C_{21}t^{-v_2}$. The dotted line corresponds to the zero value.}
\end{figure}

On the other hand, for assuring that our data analysis was properly correct, we checked the previous results by adding, to the fit formula, the leading-order scaling-corrections term. In improved models such a term ($\sim t^{-v_1}$) should be absent. We verified by fitting the data with the formula $\ln\chi(t) = \ln\chi_{0} + \rho\ln t + C_{11}t^{-v_1} + C_{21}t^{-v_2}$ and moving the fitting data range in the same intervals as we have explained in the previous paragraphs. Finally, we checked the behaviour of $C_{11}$ (see figure \ref{img_C}) and we verified his compatibility with zero value. In conclusion,  taking into account the $t^{-v_1}$ term, we did not observe any appreciably changing in the estimate of the dynamical critical exponent $z$. Furthermore, checking the fit by using the least-square test, we have not noticed any improvement by respect to the previous fits.

Furthermore, like in others works that used improved models \cite{relax_rand, Elio4}, we have took care of further corrections by checking the stability of the fits by varying the temporal range in which the fits have been done.

After that, we checked also the stability of the results by performing further simulations with different values for the parameters of the Hamiltonian. We fixed the parameter $D$ at the new value $\hat D=D^{*}+\delta D$, where $\delta D = 0.008$ is the confidence of the improved value $D^{*}=0.641$. Then, after having tuned the correct critical inverse temperature $\hat\beta_{c}$ \cite{IHT, IHT2, Vinti}, we performed Monte Carlo simulations on a lattice with length $L=96$. We have done an average over $2\cdot 10^5$ Markov chains; for each chain we set different high-temperature initial conditions and, after the quench, we have done $700$ lattice sweeps using Metropolis algorithm. Performing the same data analysis that we have done in the previous sections we obtained an estimate of the dynamical critical exponent $z=2.022(6)$. Therefore, we argued that the uncertainty on the estimate of the critical value due to the numerical uncertainty of $D^{*}$ is
\begin{equation}
\delta z \simeq |z(D^{*})-z(\hat D)|\approx 0.002(7).
\end{equation}
Although this result is compatible with zero, we have to take into account its effects on the determination of the global incertitude of the value of the dynamical critical exponent $z$. Thus, we definitively predict
\begin{equation}
z = 2.020(6)[2],
\end{equation}
where the error in brackets gives the variation of the estimate as $D^{*}$ varies within one error bar.

\subsection{Correlation Length}
\begin{figure}[t]
\center
\includegraphics[width=0.5\textwidth]{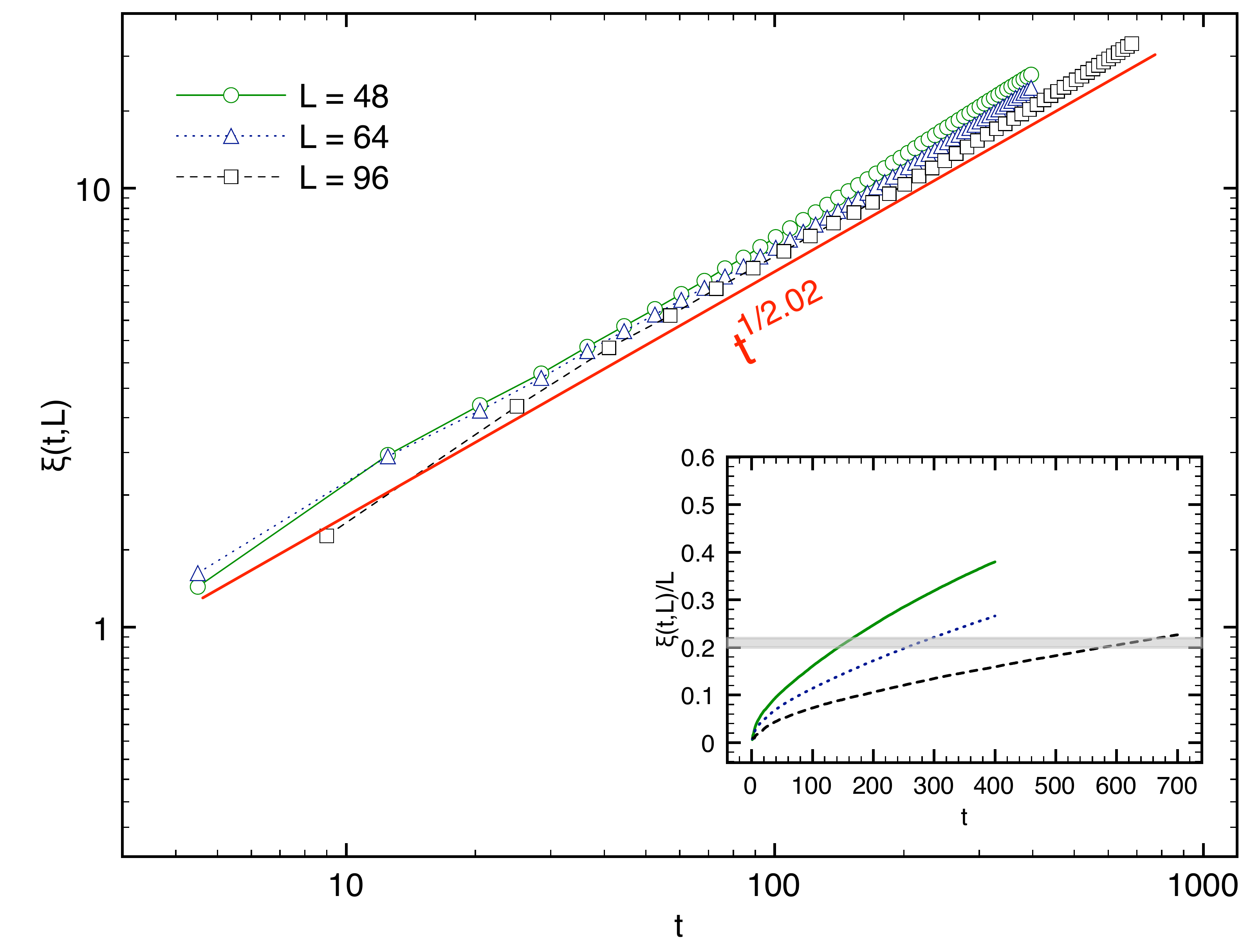}
\caption{\label{img_xi}The behaviour of the correlation length: in the main plot $\xi(t,L)$ versus $t$ in log-log scale; in the subview the rescaled $\xi(t,L)/L$.}
\end{figure}

Finally, using the correlation length, we have checked the results achieved in the last Section. In Fig.\ref{img_xi} we plot $\xi(t,L)$ and $\xi(t,L)/L$. Confronting this plot with Fig.\ref{img_ro} it is easy to see that when $\xi(t,L)/L\lesssim 0.22$ there are not finite-size effects in the behaviour of susceptibility $\chi(t,L)$. In the subview of Fig.\ref{img_xi} the gray stripe represents the data range into which susceptibility is starting to leave the infinite-volume curve. In the main graph we show the correlation length in $\log-\log$ scale; the red straight line represents the infinite-volume behaviour according to the dynamical critical exponent that we have found in the last Section.  Indeed, according to dynamical $\mathcal{RG}$, in the thermodynamic limit, correlation length behaves as
\begin{equation}
\xi(t)=\xi_{0}t^{1/z}F_{\xi}
\end{equation}
where $F_{\xi}$ is an analytic function that depends on the next-to-leading scaling fields.

As shown by Fig.\ref{img_xi}, also in this case the curves for different lattice size are parallels to the infinite-volume curve up to a value $t_{max}(L)$ which increases as $L^z$.

\section{Conclusions}
In this paper we have studied the off-equilibrium purely relaxational dynamics in 3d-Blume-Capel model. According to standard $\mathcal{RG}$ arguments applied to dynamics, the dynamical critical behaviour in such a system should belong to the same model-A dynamical universality class. If this description is correct, the dynamical critical exponent $z$ is the same as in 3d-Ising model and the leading scaling corrections are controlled by the same $\mathcal{RG}$ operators that appear in the statics and therefore are characterized by the static correction-to-scaling exponents $\omega_{1}=0.84(4)$ and $\omega_{2}=1.67(11)$. For the same reasons, in the case of improved Hamiltonians, leading scaling corrections should be also absent in dynamical quantities. Improved models are expected to provide the most precise estimates of universal dynamical quantities. For instance, in FSS studies at the critical point, corrections to scaling decay as $L^{-1.67}$ in improved models, while in generic systems the approach to the thermodynamics limit is much slower, with corrections decaying as $L^{-0.84}$. 

The main results of our analysis can be summarized as follows.

We investigate the off-equilibrium relaxational dynamics in the 3d-Blume-Capel model. We started from disordered $T=\infty$ configurations and observed the relaxation at $T=T_{c}$. The results show that previous FT estimates of dynamical critical exponent $z$ are perfectly consistent with our MC simulations that give the value $z=2.020(8)$. In the analysis particular attention has been taken to avoid finite-size corrections. 

Since the model is improved, we do not observe corrections proportional to $t^{-\omega_{1}/z}$; instead our data show corrections that are proportional to $t^{-\omega_{2}/z}$. Moreover, we have been careful at new scaling corrections introduced by a new $\mathcal{RG}$ operator that appear in the dynamics and we have shown that it is possible to forget such a corrections throughout the data analysis. Indeed, the correlations of the order parameter at initial time surface (tuned by $\tau_0^{-1}$) give rise to dynamical scaling corrections proportional to $t^{-1}$; nevertheless we can absorb such a corrections into $t^{-\omega_{2}/z}$.

In conclusion, with this work, we have found an estimate of the dynamical critical exponent $z$ for the three-dimensional Ising universality class that is perfectly in agreement with the FT predictions.

\section{Acknowledgements}
Ettore Vicari is greatly acknowledged for useful and pleasant discussions and for his continuous support during all the period in which this work was done. I am in debt with Dragi Karevski for the careful reading of the manuscript and I want to thank Pasquale Calabrese for awesome discussions. 

The MC simulations have been done at the Computer Laboratory of the Physics Departement at Pisa University.

\appendix
\section{Metropolis Algorithm}
We have implemented the standard local Metropolis algorithm with the acceptance rate
\begin{equation}
A=\min \left[ 1, \exp\left(-\beta [\mathcal{H'-H}] \right) \right]
\end{equation}
where $\mathcal{H'}$ and $\mathcal{H}$ correspond to the Hamiltonian evaluated for the proposal and for the given spin configuration, rispectively. The proposal is generated by choosing with the same probability one of the two new possible values of the spin at a single site $x$ of the lattice. For example, if the initial value of the spin at site $x$ is $+1$ then we randomly choose a new value from the set $\{0,-1\}$. Hence $\mathcal{H'-H}$ depends only on the value of the spin at the site $x$ and at its neighbours $y$. We perform the single-site update sequentially, moving from one site to next-nearest-neighbour site in a checkerboard fashion. Defining the parity of a cubic lattice site as $P(i,j,k)=(-1)^{i+j+k}$, we perform a sequential update first over all even sites then over all odd ones. The time unit is defined as a Monte Carlo sweep over all sites of the lattice. Moreover, the pseudorandom number generator is been implemented by using the 64-bit Mersenne Twister algorithm \cite{Matsumoto, Nishimura}.


\end{document}